\documentstyle[11pt]{article}
\begin{document}
\begin{center}{\bf SELF-CONSISTENT EVOLUTION OF

    RING GALAXIES }

\vspace{1.truecm}

{\bf P. Mazzei$^1$, A. Curir$^2$, C. Bonoli$^1$}
\vspace{0.7truecm}

$^1$ Osservatorio Astronomico, Padova, Italy.

$^2$ Osservatorio Astronomico, Torino, Italy.

\end{center}

\vspace{0.5truecm}

\begin{center} {Abstract}
\end{center}

Ring galaxies are commonly known as objects where a burst of star formation was
triggered by a close encounter with an intruder, maybe a satellite galaxy. BVRI
CCD observations of five ring galaxies have been performed. Here we present the
results of a self-consistent approach  to reproduce their observed morphology
and  spectral energy distribution using updated $N$--body simulations and
evolutionary population synthesis models extending from UV to far--IR
wavelengths. Some suggestions about the evolutionary properties of these
starburst galaxies are then derived.

\section{Introduction}

Although it is well known that many of the most spectacular examples of
starburst galaxies are members of interacting systems, relatively
little quantitative information is available on the overall  effects of
interactions on the star formation properties of galaxies.
The degree of enhancement in the star formation rate varies over a
large range, from galaxies which appear unaltered after the
interaction, at least when observed  at present, to galaxies with
star formation bursts 10--100 times stronger than
typically observed in isolated systems (Kennicutt, 1987; Schweitzer, 1990).
This diversified scenario
is a consequence of the large spread both in the phenomenologic
conditions of the interaction and in the dynamical and physical stage
of the colliding galaxies.
In this paper, the first of a series on interacting systems, we focus our
attention on a special class of interacting objects, the ring galaxies. They
are commonly known as spiral galaxies where a burst of star formation is
triggered by a close encounter with an intruder.
An almost normal penetration of the companion at radii up to $40\%$ of the
disc radius (Lynds and Toomre, 1976; Appleton and James, 1990) generates
their strange morphology where a more or less sharp ring surrounds an
off-centered nucleus or an empty region.
Following Appleton and Struck-Marcell (1987), rings provide both an
understanding example of extended
coherent starbursts and a laboratory study of the effects of large--scale,
nonlinear density waves on the interstellar gas.
Ring galaxies are strong
far--IR emitters. Their IR luminosity, $L_{FIR}$, as derived by IRAS
colors,  is even larger than for barred galaxies (De Jong et
al. 1984), the most luminous class of the Shapley--Ames
galaxies. Furthermore, some rings show $L_{FIR}$ also greater than the
prototype starburst galaxy $M82$. As a general rule such galaxies emit as
much infrared radiation as blue optical light. Therefore the mean value of
the blue absolute luminosity, $L_B$, of ring galaxies agrees well with that
derived by Keel et al. (1985) for Arp galaxies in general, $L_B=1.4\times
10^{10}(100/H_0)^2\,L\odot$, about 0.5 mag brighter than for non
interacting ones.

In this paper we  adopt a self--consistent approach to analyze the
morphological and photometric properties of five of them through up-to-date
$N$--body simulations and evolutionary population synthesis (EPS) models
extending from UV to far-IR wavelengths. We use the results of our $N$--body
simulations as inputs of our EPS models to derive some suggestions for modeling
the burst and the luminosities involved in the optical and far-IR ranges.
Models obtained with a more refined SPH code including an internally consistent
luminosity evolution will be treated in a later paper.

Galaxies were  selected from the list of Appleton and Struck-Marcell (1987) and
observed by one of us (C.B.) in BVRI. Far-IR (FIR) data come from IRAS
catalogue
(Version 2).

The plan of this paper is the following:
observations and data reduction are described in Section 2; corresponding
simulations of collisions between stellar disks and intruders are
in Section 3 while in Section 4 we discuss the treatment of the burst;
some suggestion is also given about the postburst evolution of such systems.
In section 5 we point out the results for each galaxy and in section 6  our
conclusions.

\section{Observations}

BVRI CCD observations were carried out at Padova-Asiago Observatory using a GEC
CCD with a pixel size of 22$\,\mu$m, corresponding to 0.28 arcsec. Exposure
times were 50m in B, 30m in V,  20m both in R and in I. Typical seeing during
the observing run was $1".8$

The reduction of images consisted in the standard procedure. Pedestal was
removed by subtracting a mean value estimated in the overscan region of each
frame. Several flat field exposures, obtained at twilight in each color, were
used to remove variations in pixel-to-pixel sensitivity. Sky background was
determined in each frame by making a histogram of the gray levels of all pixels
and locating the peak value by a gaussian fit of the low level part.

Frames of selected stars in M67 cluster, obtained in the same nights,
were used to calibrate the instrument photometry.
The uncertainty in zero points, estimated from the statistics of the
calibration constants derived from different standard stars in the same
frame, turns out to be some hundredth of magnitudes in the four bands.
However, we observed several standard fields in different nights and we
found that the measurements show rms $\simeq 0.08$. This increased
uncertainty is likely due to the extinction corrections and changes in
photometric conditions from night to night. All in all we have assumed that
the calibration error is $\simeq 0.1$ mag.

Figure 1 shows R--band isophotes of our sample and Table 1 presents their
BVRI magnitudes in the  Johnson (1966) system.

\section{N--body simulations}

We performed numerical simulations of collisions between stellar disks embedded
in static halos and suitable intruders. The code used is the Hernquist (1987)
TREECODE which employs a tree structure to calculate the gravitational
forces; this is a useful tool to study systems endowed with strong deviation
from spherical symmetry (Curir, Diaferio, De Felice, 1993).

The disks have been  relaxed down by solving numerically the Laplace equation
in cylindrical coordinates (Binney and Tremaine, 1987) and then immersed in a
massive halo structure (King model, 1981).  The system is evolved by several
rotation periods to test its  stability. The halo has an important heating
effect on the disk during the assessment.

The companions used as intruders are massive points or King spheres of
different radii. A series of central collisions have been performed varying the
angle of incidence and the velocity of the companion. We refer for a more
complete and detailed set of simulations in the space of the numerical
parameters to Curir and Filippi (1994). Here we present only the more suitable
simulations in order to match the morphologies of our selected ring galaxies.
{}From the pioneering numerical work of Lynds and Toomre (1976) many  papers
are
now available in literature on this subject. Recently, models have been
produced using  non dissipative $N$--body codes by Luban-Lotan and
Struck-Marcell (1989) and by Huang and Stewart (1988); furthermore, gas
dynamics and dissipative phenomena have been taken into account by Gerber, Lamb
and Balsara (1992), Weil  and Hernquist (1993), Struck-Marcell and Higdon
(1993) which presented very refined simulations, devoted to provide good models
for specific astronomical objects. In this paper we will use pure $N$--body
simulations to have more insights in the role played by the dynamical friction
in the case of the intrusion.

In the following we briefly summarize the main points of our model and then
present our results.

\subsection {The numerical methods}

In the system of units employed, $G=1$ ($G$ is the gravitational constant), the
length scale is 0.1, and the mass of the disk and of the intruder are equal to
1. Translating these values into physical units,  mass unit is $5 \times
10^{10} \ M_{\odot}$,  distance unit is $20\,$ Kpc and  time unit is $ 158\,$
Myr. A fixed timestep $\delta t$ of $0.01$ (time units) is used to update the
gravitational forces which are calculated including quadrupole moments of the
gravitational potential and using a tolerance parameter $\theta = 0.8$.

The  target galaxy model consists of two components: a spherical halo and an
exponential disk. The mass ratio of these components is $4:1$. The model
mass-points , tracers  of the true mass distribution, interact through the
pseudo-keplerian softened potential
$$
\Phi={{-K}\over{\left(r^2+\varepsilon^2\right)^{1\over2}}}
$$
where $\varepsilon$ is the softening parameter adopted to avoid extremely
close encounters.

The softening length for the gravitational force depends on the type
of particle: the stellar particles employ a softening $\varepsilon_d=0.015$
while halo particles employ $\varepsilon_h=0.038$

The disk and the halo are modeled as $N$--body systems with $3192$ and $5000$
particles respectively; these numbers are large enough to adequately describe
their kinematical properties (see Curir and Filippi, 1994 for a detailed
discussion on this subject).
The halo particles are assumed to be dark matter particles. The higher
value of the softening length accounts for the weakly interacting nature
supposed for these particles.

The spatial distribution of stars in the disk follows the planar surface
density law
$$
\rho(r)={\rho_0}{e^{-{{r}\over{r_0}}}}
$$
where $r_0$ is the disk scale length, equal to 2 Kpc.

To obtain each particle position-vector according to the assumed mass
density we use the technique called ``rejection method'' for generating random
deviates whose distribution function is known (Press, Flannery, Teukolsky,
Vetterling 1990). Azimuthal angles are chosen randomly in the interval
$[0,2\pi]$. The vertical co-ordinate is extracted from a gaussian distribution
with a dispersion equal to the $1\%$ of the disk scale length.

We assign  circular velocities analytically to disk stars, through a particular
expression of the potential obtained with the aid of Jeans equations. Radial,
azimuthal and vertical velocity dispersions are locally imposed on disk
particles; dispersions are monitored through a Toomre-like parameter $Q$
(Toomre, 1978), a true
"stability thermometer" for the exponential disk (in our simulations $Q=1.6$).
The $Q$ parameter is an input parameter for our model.

The halo is a King model with the characteristic distribution function:
$$
f_K(E)=\cases {{\rho {(2\pi {\sigma}^2)}^{-{{3}\over{2}}}
(e^{{{E}\over{\sigma^2}}}-1)} &when $E>0$; \cr 0 &when $E\le 0$. \cr}
$$
where  $\rho$ is the density, $\sigma $
the velocity dispersion and E the total energy of the stars.
King models are dynamically stable systems.

To obtain the final galaxy model we relax the two components together,
superimposing the gravitational potential generated from the exponential disk
on
the potential of the spherical model which has a radius twice that of the disk
{}.
The two components have approximately the same mass within three disk scale
lengths.
The resulting rotation curve of the disk is substantially flat and reproduces
quite faithfully the observed curves for a large part of spiral galaxies
(Rubin et al. 1980) (see for ex. Fig.5a, curve labeled 1).

The energy is conserved better than $0.5 \%$ and the angular momentum better
than $0.01 \%$ over the whole run for each model.

\subsection { Results}

We performed several simulations using different intruders and a different
impact velocity, defined as the initial relative velocity between the two
systems placed 60 Kpc a part. The mass of the intruder is equal to the mass
of the disk target (Curir and Filippi, 1994). In our space of parameters we
singled out  two different regimes for the ring formation.

In the first the impact velocity is below the escape velocity of the intruder (
i.e $\le 200\, km\, s^{-1}$) and the dynamical friction dominates
the interaction.
As the intruder oscillates crossing the disc, it loses a remarkable part of its
mass  and is finally trapped in the system. The time evolution of the ring
structure is $190$ millions of years depending slowly on the radius and on the
velocity of the companion.
The ring appears while the intruder is very near to the disk, between $0.$ and
$1.5$ in our  numerical units (i.e. 30 kpc).
Fig.s 2a and 2b show an edge--on and the face--on view respectively of the
predicted evolution for a typical case in this regime. This is characterized by
a range of velocities from $\approx 100\, km\, s^{-1}$ to $200\,km\, s^{-1}$.

In the second regime
the dynamical friction is much less active because the impact
velocity is higher than the escape velocity, thus the ring is
formed when the intruder is already far from the disk, at a distance  between
$1.5$ (30 Kpc) and $3$ (60 Kpc).  The intruder emerges from the collision very
unperturbed in shape but enlarged in volume (on the average the radius is
almost twice the original one). The time evolution of the ring structure is
$112$ millions of years when the radius of the intruder is equal to
0.28 of the target disk.
As far as these high impact velocity cases are concerned,
Fig. 3 shows the time evolution of the morphology related to the formation of a
ring with a nucleus (the projected intruder) whereas Fig. 4 describes an empty
ring.
In Fig. 5 the behaviour of the rotation velocity (panel a)) and that of its
components in the polar (panel b)) and in the radial (panel c)) directions, for
the same simulation as in Fig. 3, is presented; in particular the curve number
4 of panel b) shows that  the disk  relaxes to a relatively high value of the z
component of the velocity dispersion after the merger.

By performing different simulations using smaller and smaller companions we
obtain practically the same configurations provided we scale the impact
velocity in a linear way with the radius of the intruder. However, the time of
evolution, $\Delta t$, lengthens. In particular a simple material point behaves
in this regime like a King sphere (of unit radius) giving rise to a ring living
$350$ millions of years.
So in this regime we can define the stage of the evolution with a
parameter $\tau$, given by the ratio $t/\Delta t$ where t starts with the
beginning of the interaction.
We point out that the two regimes previously outlined are not the only two
possibilities for the formation of a ring galaxy. There are more other
possibilities which we did not analyze, like the disruption of a small
companion, which may generate an empty ring, or the presence of a nucleus
inside the ring simply because the target retains its bulge, which we did not
include in our simulations; Curir and Filippi (1994)  explored also the
possibility of a ``spontaneous" ring formation.  In particular  these regimes,
driven by the dynamical friction, are simply related to the mechanism of the
intrusion and provide good numerical fits for our sample of ring galaxies.

The related density waves have quite different features (Fig. 6).
In the first one the wave appears as a light perturbation of a decreasing
exponential density distribution.   At the same time its amplitude is much less
than in the second regime, where the wave expands starting from a strong
central depression. In both cases, well after the collision,  the system
settles into an exponential disk thicker and larger than the original one.

\section{Photometric properties from UV to far--IR: their evolution}

Our EPS models, accounting for the dust effects, provide
chemical and photometric evolution of a galaxy in a self-consistent way
from ultraviolet up to 1 mm.
The synthetic SED incorporates stellar emission, internal extinction and
re-emission by dust. The stellar contribution, including all evolutionary
phases of stars born with different metallicities, extends as far as
25$\,\mu$m.
Dust includes a cold component, heated by the general  radiation field,  and a
warm component associated with HII regions.
Observations in the short wavelength range (25--60$\,\mu$m) and in the long one
(over 60$\,\mu$m) allow us to disentangle their luminosities and temperatures.

Emission from policyclic aromatic hydrocarbon molecules (PAH) and from
circumstellar dust shells are also taken into account; the bulk of their
emission centers in the 5-25$\,\mu$m spectral range, where the pure stellar
contribution rapidly fades. This model has been successfully applied to spiral
galaxies, with particular attention to our own galaxy (Mazzei,Xu and De Zotti,
1992, hereafter referred to as MXD92) and to early-type galaxies (Mazzei and De
Zotti, 1994a; Mazzei, De Zotti and Xu, 1994).
In Appendix we briefly summarize the main points of this model.

We remember that the star formation rate (SFR) and the initial mass function
(IMF) together with its lower and upper mass limits, are important model
parameters. As discussed  in MXD92, the metallicity of the system depends
on their values
and, as a consequence, the far--IR output. Since we are modeling in a
self--consistent way both the chemical and the luminosity evolution, given the
IMF with its mass limits, the metal enrichment and the residual gas fraction,
as well as the total far--IR luminosity, are only dependent on the assumed SFR.
This is slowly decreasing before the interaction since, due to the previous
discussion (Sect. 3), a disk-like configuration  well represents the
unperturbed state. However a central collision with a spherical intruder may
trigger a strong burst of star formation (Kennicutt et al. 1987; Wright et al.
1988) lasting some $10^8$ years (see Sect. 3).
Therefore we modify the normal disk evolution (see also MXD92) superimposing a
strong burst of star formation at an age of $12\,$Gyr.

In this paper we present the first effort to analyze  the effect of a burst on
the overall SED of galaxies.

We follow a general approach without any change of the IMF as well as
of its lower and upper mass limits during the burst; this burst
is defined by its
length and intensity, which is the ratio between the value of the SFR at its
beginning and that before its onset.

Fig. 7 shows the evolution of the gas fraction, metallicity and
optical depth (see Appendix), for models computed with different $m_l$ values
and different burst length; in Fig. 8 their color evolution is presented. We
emphasize that the burst affects only the UV--near IR SED whereas the
far--IR output, depending on the optical depth, is a strong function of
$m_l$ value alone.

The shape of the far--IR SED depends on three parameters, $I_0$, $I_w$ and
$R_{w/c}$ which represent the maximum value of the interstellar radiation
field, the average energy density inside HII regions and the warm to cold
luminosity ratio respectively  (see Appendix for more details). These are
completely defined by fitting the observed IRAS colors. Given the strong
internal consistency of our model, there is not a large range of possible
values for these far--IR parameters. As we will discuss with more detail,
we suggest two different combinations of such parameters able to match the
{\it  overall} SED of a galaxy, consistent with
our $N$--body simulations. In the following, in fact  we discuss the
treatment of the burst and the post--burst phases as derived from a
self-consistent approach with $N$--body simulations.
These simulations provide the length of the burst and the spatial distribution
of the matter inside our galaxies. The first one allows  us to reduce the
number of free parameters needed to describe the burst whereas the second one
gives us some information on the physical conditions which are affecting the
shape of the far--IR SED.

\subsection{The burst phase.}

As far as optical and near-IR data only are concerned, the parameters defining
the burst, its length and intensity, are not independent but inversely
correlated. UV data, coupled with near--IR observations, could in
principle disentangle their effects. In fact, the stronger the burst, the
larger the number of massive stars, i.e. the UV flux, provided from the burst.
Thus blue UV-V, V-K  colors could suggest stronger and younger bursts, whereas
red colors weaker and older stages since red supergiant stars, AGB stars, will
appear  $\approx 10^8\,yr$ after
its beginning.
Nevertheless the extinction smooths away these differences making the galaxies
appear older and  the SFR lower. However the energy absorbed by dust in the
short wavelength region of the galaxy spectrum must be re-emitted  in the long
wavelength range. So in principle the observed overall  SED entails all the
information necessary to understand the behaviour of the SFR and to define the
model parameters.

Ring galaxies are very faint objects so UV observations are extremely rare;
only VV787 has been observed (Schultz et al. 1991) and its data suffer from
aperture correction problem.
It is well known, however, that the far--IR output of starburst galaxies is
higher than that of normal galaxies (Soifer et al. 1987; Joseph, 1990). Ring
galaxies also match these expectations (Appleton and Struck-Marcell, 1987):
some of them in fact have been detected by IRAS satellite.
Our model, by means of its  large spectral coverage, provides a useful tool for
understanding the available data.

Taking into account the previous considerations we use the time spent in the
interaction suggested by $N$--body simulations as an input for our EPS model:
the lifetime of the ring defines the length of the burst.
Results are not strongly dependent on the behaviour of the SFR during the burst
if we adopt a SFR constant or quickly decreasing with the same dependence on
the mass of gas as before the onset of the burst.

Given the burst parameters, the far--IR output of the system strongly depends
on the lower mass limit of the Salpeter IMF, $m_l$.
The higher the value of $m_l$, the larger the number of massive stars formed
and then the larger the amount of heavy elements
provided;  this enhances the extinction effects simply because the optical
depth
of the system is proportional to the metallicity of the gas (see Fig. 7).
We find that $0.01 \le m_l\, (m\odot) \le 0.20$ accounts for the overall
properties of our sample. Most ring galaxies are characterized by $L_{FIR}/L_B$
ratios larger than those of normal galaxies (Appleton and Struck-Marcell,
1987),
thus, following from our approach {\it the largest $L_{FIR}/L_B$ ratios suggest
the highest $m_l$ values}.
It is not surprising that ring far--IR SED may be  different from that of a
``normal'' spiral since the interaction greatly perturbs the initial
distributions of gas and dust.

We attempt an approach consistent with the results of $N$--body simulations
(cfr. 3.2) also in modeling the far--IR emission, in particular as far as the
cold dust distribution is concerned. The luminosity of  such a component, in
fact, depends on the interstellar radiation field distribution through the
parameter  $I_0$, i.e. the central value of its energy density (see Appendix
for more details). As discussed in the previous section,  strong collisions
provide high compression and concentration of matter in the density wave. At
the same time a large depletion of gas and stars is produced in the central
regions. According to Leisawitz and Hauser (1988), a remarkable  fraction of
the luminosity of OB stars can escape from Galactic HII regions and may
contribute to the diffuse energy density, depending both on the efficiency of
the star formation and on the optical depth in the ring (i.e. on the local gas
properties in the star forming regions). So, we expect to match the IRAS colors
using two very different values of the maximum radiation field, $I_{r_0}$ or
$I_{0}$.
We define $I_{r_0}$ as the maximum amplitude of a diffuse radiation field
centered in the density wave instead of in the disk ($I_0$). Following the
previous discussion we expect the  values of $I_{r_0}$ larger than $I_0$.
These possibilities, i.e. the "$I_{r_0}$" and the "$I_0$" cases,
correspond to different physical conditions inside the galaxies, of course, in
particular different amount of cold dust, lower for $I_{r_0}$ than for $I_{0}$,
is provided.

Given the strong internal consistency of our model,  different combinations of
the far--IR parameters with the same $m_l$ value are ruled out.
The far--IR distribution extends, indeed, to the longer wavelength the lower
$I_0$ or $I_{r_0}$. Therefore, for an observed $L_{100}/L_B$ ratio, the
decrease of $I_0$ or $I_{r_0}$, for example, requires higher far--IR output to
match the same IRAS colors, thus higher $m_l$ values.
In the following section we will present with more details the results of
models describing two extreme physical situations compatible both with
observations and with our simulations i.e. the warmest $I_{r_0}$ value,
corresponding to the lowest $m_l$, and the coldest $I_0$ case, corresponding to
the largest $m_l$ (see Tables 2 and 3).
Models computed with $m_l$ values inside this range could also match the data;
in particular rising $m_l$ we have to lower $I_{r_0}$, to increase the
$R_{w/c}$ ratio and may be to decrease the $I_w$ value.

As suggested from Fig. 9, {\it submillimeter observations could provide a
useful test to discriminate} between these possibilities keeping in suspence
the global far--IR luminosity and the dust content of such galaxies.

In Fig. 9  the overall synthetic SEDs for our sample galaxies are presented;
heavy lines show the fit obtained in the "$I_{r_0}$ case"  (see Table 2) light
lines are in the opposite one, corresponding to the "$I_0$ case" (Table 3);
for comparison, MXD92 found for the maximum radiation field a value of
$7I_{local}$.

\subsection{The postburst phase.}

For deriving the overall synthetic SED some assumption concerning the shape of
the far--IR SED (i.e. the values of the far--IR parameters) is needed, since at
this stage no observational constraints on the far--IR appearance of postburst
galaxies, faint FIR emitters, of course, are available.
Therefore  we can extract some useful information from our
$N$--body simulations. A more realistic picture will be available in the next
future  with the help both of more sensitive far--IR measures, provided from
ISO
satellite, for example, and of high resolution images (HST) of galaxies. These
would be  compared with the results of dynamical ($N$--body) and hydrodynamical
simulations (SPH) to identify the post--burst cases.

As discussed in Sect. 3.2 the final effect of the interaction is to decrease
the diffuse energy density with respect to its initial unperturbed value since
the system re-arranges in a disk thicker and larger than the unperturbed one.
We come to the same conclusion with an independent approach, i.e. taking into
account that the number of OB stars, which may strongly contribute to the
diffuse radiation field, in particular in the ``$I_{r_0}$ case", rapidly fades
after the burst. Thus we can estimate the far--IR emission of the cold dust
component, long after the burst, simply assuming $I_{r_0}$ proportional to some
power, $\alpha$, of the number of OB stars. Models suggest $\alpha \approx
0.3-0.4$.
The same suggestion cannot apply, of course, in the `` $I_0$ case"  since OB
stars  do not provide a substantial feeding to diffuse radiation field which
is,
in fact, much lower than in ``normal'' disks also during the burst.
The luminosity of the warm emission can be derived assuming the same $R_{w/c}$
ratio as for nearby spirals ($R_{w/c}=0.43$, Xu and De~Zotti, 1989),  $I_w$
being the same as during the burst.

Fig.s 10(a-e) compare the SEDs of our galaxy sample during the burst
(heavy line) and $2\,$Gyr after the burst (light line) i.e at an age of
$14\,$Gyr.

$L_{FIR}/L_B$ ratios as large as 10 times the galactic one (of about $2.5$,
MXD92) can arise also at $14\,$Gyr, depending on the $m_l$ value needed to fit
the observed data.

The bolometric luminosity of the systems is approximately 10 times lower than
during the burst and the B luminosity drops by a factor of approximately
10--30. Thus these systems will appear as very faint red galaxies. The past
interaction, which exhausted a significant amount of the residual gas
($35\%-60\%$ for our sample), reflects in the optical region of the spectrum
through very red $B-V$ colors, redder than the  average ones of normal disk
galaxies.
At $14\,$Gyr  models suggest $0.85\le B-V \le 1$, where the reddest colors
correspond to the greatest $m_l$ values (see also Fig. 8). At the same age
$V-K$ ranges from about $3.5\,$mag, as for an unperturbed disk, up to $4\,$mag
and $4.4\,$mag for $m_l=0.01\,,0.05$ and $0.1\,m\odot$ respectively. Only
$NGC\,2793$ deviates from this behaviour. Its colors are not strongly affected
since the burst consumed a low fraction of the residual gas. Therefore this
appears as a very blue galaxy, its colors being typical of an Irr also long
after the burst.

\section{Comparison with observations }

In this section we attempt a careful description of the morphological and
photometric properties of each galaxy in our sample. The observed isophotes
(Fig. 1) will be compared with  suitable isodensity contour levels obtained
from $N$-body simulations (Fig. 11). The particle positions of the numerical
dump chosen to represent the galaxy are projected on a suitable plane and then
the isodensity contours are extracted. The time spent in the interaction, as
described in Section 3, is the length of the burst for our EPS model.
This length is practically constant in the first regime (0.2 Gyr) whereas it
depends on the radius and on the velocity of the intruder in the second one,
ranging between 0.11 and 0.35 Gyr. In this case we refer our fits to the
parameter $\tau$ which settles the stage of the burst evolution (see Sect.
3.2);
it is not surprising that such a stage may be different for our morphological
and photometric fits although, in most the cases, they well agree.

We derive some suggestions about the strength of the collision or the intensity
of the burst (as defined in Sect. 4), the total far--IR emission,
$L_{FIR_{tot}}$, the absolute bolometric luminosity, $L_{bol}$,  and the
residual amount of gas.

We point out that, given the internal consistency of UV--optical and far--IR
SEDs, widely discussed both in Sect. 4 and in the Appendix, only two
combinations of $m_l$ and far--IR parameters for each galaxy are discussed.
These correspond to the extreme ``$I_{r_0}$" and ``$I_0$" cases respectively
(see Table 2 and 3). Submillimeter observations should be careful recommended
to reduce the uncertainty further.

For reasons of comparison we summarize here some useful values derived for our
own galaxy (MXD92): $L_{FIR_{tot}}/L_B=2.6$, the corresponding ratio using
$L_{FIR}$, i.e. the far--IR luminosity computed  from $42.5$ to $122.5\, \mu
m$, $L_{FIR}/L_B\simeq 0.9$, and $L_{FIR_{tot}}/L_{bol}\simeq 0.3$.

We used $H_0=50\,km\,s^{-1}\,Mpc^{-1}$.

\subsection {VV 789  or I  Zw 45}

The isodensity contour levels are represented in Fig 11a (for comparison see
Fig. 1). The simulation describes a collision of a disk target lying in the xy
plane with a King sphere having a radius 0.28 times that of the disk. The
velocity of the intruder is $1140\,km\,s^{-1}$ inclined of $30\deg$ with
respect to the normal, z, to the disk plane. The stage of the ring evolution
correspond to $\tau=0.20$.

The overall SED of this galaxy (Fig. 9(a)) requires  a strong burst of star
formation whose intensity being 60. Models have been computed with a lower mass
limit of $0.01$ and $0.05\ m\odot$ respectively for the $I_{r_0}$ (heavy line)
and $I_0$  (light line) cases. Our photometric fits correspond  to different
stages of the burst evolution, $\tau=$0.75 and 0.28 respectively.

The residual fractions of gas are $0.25$ and $0.54$ corresponding to
(1.56--4.07)$\, \times 10^{10}\ m\odot$ respectively. For comparison Theys and
Spiegel (1976) derived a value of $4.56\times 10^9\ m\odot$ for the mass of
neutral hydrogen alone. The system appears as a very blue luminous  galaxy with
$L_B\,(10^{10}\ L\odot)\simeq 4.5$ (Appleton and Struck--Marcell, 1989).

{}From the previous models we derive $L_{FIR_{tot}}/L_{bol}=$0.50 and 0.68,
which
correspond to $L_{FIR}/L_{bol}=$0.44 and 0.21. Moreover we predict
$L_{FIR}/L_B=2.01$ and 1.59 respectively; the last ratio increases to 2.3 and
5.24 taking into account the total far--IR luminosity; thus the long
wavelength range could include up to $70\%$ of the far--IR luminosity of this
galaxy.
\medskip

\subsection {VV 330 or UGC 5600}

The parameters describing the intrusion are the same as
for VV798 but the collision velocity is orthogonal to the plane of the disk.
In Fig. 11b we present the morphology of the simulated system (for
comparison see Fig. 1) taken to a stage $\tau=0.40$ of the ring evolution.

Two extreme  models, computed with different lower mass limit of the IMF,
$0.05$ and $0.1\ m\odot$ respectively, but the same intensity of the burst, 60,
match quite well the optical SED and the IRAS colors of this galaxy (Fig. 9b).
Our fits correspond  to similar stages of the interaction, $\tau=0.43$ and
$0.29$ respectively, where the residual fractions of gas are $0.47$ and $0.63$
respectively.
We derive a blue luminosity, $L_B=0.56\times 10^{10}\ L\odot$, like normal
galaxies (De Jong et a. 1984) however a large fraction of the bolometric
luminosity comes out at the longest wavelengths:
$L_{FIR_{tot}}/L_{bol}=$0.70--0.68 respectively. We find $L_{FIR}/L_B=$3.12 and
3.04 which rise up to 6.0 and 7.6 including the total far--IR luminosity.
\medskip

\subsection{VV 787 or Arp 147}

In a recent paper Gerber et al. (1992) showed that this galaxy is  the result
of
an off--center high velocity collision between an elliptical and a spiral of
the
same mass, the intrusion being perpendicular to a disk plane. This corresponds
to a collision with a impact velocity $v\simeq 450\,km\,s^{-1}$ which well
agrees with our second velocity regime. Their isophote fit suggests $tau \simeq
0.66$ from the beginning. For the same stage, models matching quite well
the optical SED  and the IRAS colors of this galaxy (Fig. 9c) correspond to
$m_l$ $0.05$ and $0.1\ m\odot$ with intensities of 60 and 100 respectively.

The blue luminosity of this galaxy is about $1.4 \times 10^{10}\, L\odot$
(Appleton and Struck--Marcell, 1989). We derive large $L_{FIR_{tot}}/L_{bol}$
ratios, 0.71--0.78 respectively, $L_{FIR}/L_B$ ratios of 2.70 and 4.50,
increasing of the same factor, 2.5, including the global far--IR emission, and
residual fractions of gas of 0.54 and 0.43 respectively.
\medskip

\subsection {VV 32 or Arp 148}

The impact velocity used for this simulation is $160\,km s^{-1}$. Thus this
ring is the only one in our sample described by the first velocity  regime (see
Sect. 3.2). This result confirms previous optical and mid-infrared analyses
(Joy and Harvey, 1987) suggesting that this galaxy is coalescing. In Fig. 2 we
have shown the time evolution of its morphology. Fig. 11c presents the
corresponding isodensity curves (for comparison see Fig. 1).

The BVRI region of the SED of this galaxy has been analyzed by Bonoli  (1987)
and we refer to that paper for a detailed discussion. Models computed with the
same burst intensity, $i=60$, and different mass limits, $0.1\,m\odot$  and
$0.2\,m\odot$, match well the overall SED of such a galaxy (Fig. 9d) for a
warm and a cold diffuse radiation field respectively (see Tables 2 and 3). The
$L_{FIR}/L_B$ ratio suggested by these models, $6$ and $7$ respectively, one of
the greatest in
the sample of ring galaxies by Appleton and Struck-Marcell (1987),
becomes  $10$ and $16$ including the total far--IR luminosity.
All models predict that dust absorbs
about $80\%$ of the bolometric luminosity
of the galaxy.
Even if the the burst consumes $75\%$ and $40\%$ of the residual mass of gas,
the system retains a substantial gas fraction, $\approx 0.4$ in both the cases
which correspond to different stages of the burst evolution, 0.15 and 0.05 Gyr
from its beginning respectively.

\subsection {$NGC\,2793$}

Its morphology is obtained with an head--on collision of a material point. The
impact velocity is $600\, km s^{-1}$. The simulated isodensity contours of the
system seen face--on are represented in Fig. 11d at $\tau=0.17$.

Our results suggest a very early stage in the burst development, in agreement
with the short radius observed for this ring (Appleton and Struck-Marcell,
1989). Fig.9e, heavy line, shows that the overall IRAS SED of this galaxy is
matched well by a very high radiation field, $I_{r_0}=45I_{loc}$ (see Table 2),
as before the onset of the active phase of star formation in the ring: stars
compressed
in the ring increase the diffuse energy density, starting then  the new
process of star formation triggered by shocks. Models suggest:
$L_{FIR}/L_{FIR_{tot}}=0.5$ and $L_{FIR}/L_{bol}=0.22$.

At
the onset of the burst, whose intensity is 10, a low central radiation
field and a large amount of warm dust,  about 3 times larger than in our own
galaxy, are required to match the data (see Table 3).
The far-IR output is leaded by the warm component which entails
$40\%$ of the total IR luminosity of the galaxy instead of $17\%$ as in our own
galaxy; moreover, although the ratio $L_{FIR}/L_{FIR_{tot}}$ is practically
the same as in our own galaxy,
$0.44$ and $0.39$ respectively, the warm emission encompasses about $65\%$ of
$L_{FIR}$ instead of $14\%$.
The $L_{FIR}/L_B$ ratio is approximately double of the galactic one and
$L_{FIR_{tot}}/L_B$ is larger by a factor of 1.5.
However $L_{FIR_{tot}}/L_{bol}=0.4$, like the Galaxy.

The lower mass limit for the best fit model is $0.01\ m\odot$.

We derive a large fraction of residual gas, $0.65$, and an absolute blue
luminosity $L_B=3.26\times 10^9\ L\odot$, the lowest in our sample. Our fit
implies a very early stage in the burst evolution corresponding to
$\tau=0.05$ and to a residual gas consumption of only $15\%$. The burst does
not affect the colors of this system assuming the typical colors of an Irr
during whole evolution.
\medskip

\section{Conclusions}

We investigate the evolution of ring galaxies in a self-consistent way
using up-to-date $N$--body simulations and EPS  models providing
the overall SED, from 0.06$\,\mu$m up to 1 millimeter .

Results are compared with optical and far--IR data of a sample of 5 ring
galaxies selected from the Appleton and Struck-Marcell (1987) list. Data from B
to I bands are derived from our CCD photometry, far--IR data come from IRAS
catalogue (Version 2).

Although in principle a ring galaxy can arise from a large number of
situations (Curir and Filippi, 1994 and references therein), nevertheless the
morphologies of our sample are well reproduced by a central collisions between
a
disk galaxy and a spherical intruder. We singled out two different behaviours
for the ring formation, according with the collision velocity being below or up
the escape velocity. In the former one the ring seems to be a transient
structure, ending in a complete merging. VV32 is an example of this regime
leaded by the effect of dynamical friction and lasting about $0.2\, Gyr$.

Stronger collisions, instead, producing an empty ring (i.e VV787), require
higher impact velocities. In this second regime the length of the
interaction depends on the radius of the colliding galaxy: the smaller the
radius the longer the interaction providing a ring configuration. When the
burst turns off, after $0.1-0.35\,$Gyr from its beginning, a new disk-like
configuration would arise owing to the re-arrangement of the galaxy. This will
appear as a disk with a slightly larger radius  and a higher thickness.
These simulations give insights in the modeling the photometric properties of
these objects providing the length of the burst and the distribution of the
matter inside our galaxies. The former one enables us to reduce the number of
free parameters needed to describe the burst whereas the second one suggests
two different extreme values of the radiation field inside these galaxies both
matching well their IRAS colors.
It is well known, indeed, that rings are powerful far--IR emitters (Appleton
and
Struck--Marcell, 1987), furthermore their dust temperature, as suggested by
their $f_{60}/f_{100}$ ratio, is  also warmer than isolated disk galaxies. We
find  $L_{FIR}/L_{bol}$ ratios exceeding those of normal galaxies by at least a
factor of two, a value of 0.7 being typical. As suggested from our EPS models,
the lower mass limit of the IMF is the most important parameter driving the
far--IR output of these systems. Values up to 20 times greater than those
derived to match the overall SED of our own galaxy ($m_l=0.01$, MXD92) are
needed.
{}From our synthetic SEDs we derive $L_{FIR_{tot}} \ge 2 L_{FIR}$ being
$L_{FIR}$ computed from 42.5 and 122.5$\,\mu$m, with a warm luminosity which
entails up to one half of the whole far--IR emission.
Unfortunately their global far--IR luminosity is uncertain for the lack
of submillimeter data. Observations in this spectral domain will provide useful
information on the distribution and the amount of dust in such galaxies.
Data in the mid--far IR would also provide important hints on the understanding
the nature of the dust in ring galaxies.
Observations with ISO satellite, whose launch is scheduled for september 19,
1995 will be performed. ISO instrumentation, with its high sensitivity,
better than IRAS, and its larger spectral coverage, will allow us to better
define the observed SEDs from 7 up to 200$\,\mu$m.
Observations in the short wavelength range will provide useful information on
PAH and warm dust contributions, those in the long wavelength one, in
particular at 200$\,\mu$m, will enable us to disentangle models with different
maximum radiation field solving the puzzle of cold dust amount and
distribution as so as of the true $L_{FIR}$ luminosity.

After the burst the star formation rate lowers to a quasi-constant value
driving
a very slow system evolution.
In time the bolometric luminosity of the galaxy will be dominated by a
number of red giant stars larger than in normal galaxies.
Systems which have experienced a ring phase characterized by a strong burst
of star formation,  would keep a larger
$L_{FIR}/L_B$ ratio than normal disk galaxies as a consequence of the past
interaction.
The predicted SED  will reflect this situation showing very red optical
near--IR colors and a slightly cooler far--IR emission
following the re-arrangement of the disk.
Systems which have experienced a burst of low intensity, like $NGC\,2793$,
will appear as very blue low brightness galaxies.

\section{Appendix}

In the following we summarize the fundamental assumptions of the model which
allows us to derive the SED of galaxies over the whole frequency range, from UV
($\lambda=0.06$ $\mu\,m$) to far-IR ($\lambda=1000$ $\mu\,m$) ( see MXD92 for
more details).
\medskip

\subsection {The chemical evolution model}

We have adopted a Schmidt (1959) parametrization, wherein the
star--formation rate (SFR), $\psi (t)$, is proportional
to some power of
the fractional mass of gas in the galaxy, $f_g = m_{gas}/m_{gal}$,
assumed to be, initially, unity ($m_{gal}=10^{11}~m\odot$).
$$\psi (t) = \psi _0 f_g^n\, m\odot\,yr^{-1}. \eqno(1)$$
The initial mass function (IMF), $\phi (m)$, has a Salpeter (1955) form:
$$\phi (m) dm = A \left({m\over m\odot}\right)^{-2.35} d\left({m\over
m\odot}\right)\qquad m_l \leq m \leq m_u, \eqno(2)$$
with $m_u=100\,m\odot$ and $m_l\le 0.2\,m\odot$ (see text).

The influence of a different choice of the power law
index, n, for the dependence of the SFR on the gas density has been discussed
by Mazzei (1988). The effects of different choices for the IMF and its lower
mass limit, $m_l$, are analysed in MXD92
for late--type systems and in Mazzei et al. (1994) for early--type
galaxies.  The general conclusion is that the overall evolution of late--type
systems is weakly  depending on n.
We put $n=1$ and $\psi_0=4\, m\odot/yr$.

The galaxy is assumed to be a closed
system with
gas and stars well mixed and uniformly distributed.
However, we do not assume that recycling is instantaneous, i.e. stellar
lifetimes are taken into account.

The variations with galactic age of the fractional gas mass $f_g(t)$
[and, through eq. (1), of the SFR, $\psi(t)$] and of the gas
metallicity $Z_g(t)$ are obtained by numerically solving the standard equations
for the chemical evolution.

As far as the burst is concerned we point out that the observed blue and red
morphologies of our selected galaxies are well fitted by a density wave
sweeping and compressing a large fraction of the matter in the galaxy (compare
Fig. 1 with Fig. 11) which may give rise to a strong burst of star formation.
So
the  closed  box approximation models  the chemical evolution well
during the burst.

\subsection {Synthetic starlight spectrum}

The synthetic spectrum of stellar populations as a function
of the galactic age was derived from UV  to 25 $\mu\,m$.
The global luminosity at the galactic age $t$ is then obtained as the sum
of the contributions of all earlier generations, weighted by the appropriate
SFR as  described in MXD92.

The number of stars born at each galactic age $t$ and their metallicity
are obtained by solving the equations governing the chemical evolution,
with the SFR and IMF specified above.

To describe their distribution in the H--R diagram we have adopted
the theoretical isochrones derived by Bertelli {\it et al.}
(1990) for metallicities Z=0.001 and Z=0.02, extended by Mazzei (1988)
up to $100\,m\odot$ and to an age of $10^6\,$yr. Isochrones include all
evolutionary phases from the main sequence to the stage of planetary ejection
or of carbon ignition, as appropriate given the initial mass.

\subsection {Correction for internal extinction}

The internal extinction has been taken into account assuming that stars and
dust are well mixed. The optical depth depends on the dust to gas ratio that we
assumed to be proportional to a power of the metallicity, as in Guiderdoni and
Rocca--Volmerange (1987). Further details are given in MXD92.

\subsection {Emission from circumstellar dust}

The mid--IR emission from circumstellar dust shells was
assumed to be dominated by OH/IR stars ( see MXD92 for a discussion).
The spectrum of OH~27.2+0.2 (Baud {\it et al.}, 1985)
was assumed to be representative for stars of this class, then the total
luminosity of OH/IR stars in the passband  $\Delta\lambda$ has been computed as
in MXD92 (eq.[13]).

\subsection { Diffuse dust emission}

The diffuse dust emission spectrum takes into account
the contributions of two components: warm dust, located in regions
of high radiation field intensity (e.g., in the neighborhood of OB clusters)
and cold dust, heated by the general interstellar radiation field.
The temperature distribution of the warm dust has been parameterized as a
function of $I_w$, the average energy density inside HII regions.
Different values of this parameter entail a some change in the mean physical
conditions inside HII regions, i. e. in
the neutral hydrogen density, in the $m_l$ value, in the
efficiency of star formation, and so on.
The range of tested values to match the observed far--IR SED is
$20\le I_w/I_{loc} \le 200$ where $I_{loc}$ is the local value (see also Xu
and De Zotti, 1989).

The temperature distribution of cold dust depends on the
central intensity of the interstellar radiation field, $I_0$,
and on its distribution, $I(r)$, exponentially decreasing with the galactic
radius, r.  The range of tested values to match the far--IR SED is
$2\le I_0/I_{loc} \le 45$.

The model allows for a realistic grain--size distribution and includes
PAH molecules   (see Xu and De~Zotti (1989) and MXD92 for more details).
The amount of starlight absorbed and re--emitted by dust
is determined at each time using the model for internal extinction mentioned
above.

The relative contributions of the warm and cold dust components are also
evolving with galactic age, the warm/cold dust ratio, $R_{w/c}$, being
proportional to the star formation rate.
\section{Acknowledgments} Thanks are due to L. Hernquist who kindly provided us
with his TREECODE and to R. Filippi who produced some of the software used in
 the numerical simulations.
We thank also the referee, Curtis Struck--Marcell, for his useful comments to
improve our manuscript.
\section{References}
\parindent=0pt

Appleton, P.,N., Struck-Marcell, C.: 1987, Ap. J. 312, p.566

Appleton , P. N., James,R., A. in "Dynamical and Interactions of Galaxies"
Wielen R. editor, Springer Verlag Berlin, Heidelberg 1990, p. 200

Baud, B., Sargent, A.J., Werner, M.W., Bentley, A.F., 1985, A. J. 292, 628.

Bertelli, G., Betto, R., Bressan, A., Chiosi, C., Nasi, E., Vallenari, A.:
1990, A. A. S. 85, 845

Binney J., Tremaine S. in " Galactic Dynamics" , Princeton U. P. , 1987

Bonoli, C., 1987, A.A. 174, 57

Curir A., Filippi R.: 1994, A. A. submitted

Curir A. , Diaferio A., De Felice F.: 1993, Ap. J. 413, 70

de Jong, T., Clegg, P.E., Soifer, B.T., Rowan-Robinson, M., Habing, H.J.,
Houck, J.R., Aumann, H.H., \& Raimond, E. 1984, Ap. J. 278, L67

Gerber, R.A., Lamb, S.A., Balsara, D.S. 1992, Ap. J. 399, L51

Guiderdoni, B., Rocca--Volmerange, B.: 1987, A. \& A. 186, 1

Hernquist, L. : 1987, Ap. J. Suppl. 64, 715

Huang, S., Stewart, P.: 1988, A.\& A.197, 14

Johnson, H.L.: 1966, Ann. Rev. Astr. Astrophys. 4, 193

Joseph, R.D., 1990, in "Dynamics and Interactions of Galaxies",
Wielen R. editor, Springer Verlag Berlin, Heidelberg 1990, p. 132.

Joy, M., Harvey, P., 1987, Ap. J., 315, 480

Leisawitz, D., Hauser, M.G.: 1988, Ap. J. 332, 954.

Keel, W.C., Kennicutt, R.C., Hummel, E. \& van der Hulst, J.M. 1985, A.J.
90, 708

Kennicutt, R.C.jr.: 1987, A. J. 93, 1012.

King I. R. - Quarterly Journal of the R.A. S. , 22, 227 (1981)

Klein, U., Wielebinski, R., Morsi, H.W., 1988, A.A. 190, 41

Luban--Lotan,P., Struck--Marcell, C.: 1989, Ap. S., 156, 229

Lynds R., Toomre A. : 1976, Ap. J. 209, p. 382

Mazzei, P. 1988, Ph.D. thesis, Intern. School for Advanced Studies, Trieste.

Mazzei, P., Xu, C., De Zotti, G.: 1992, A. \& A. 256, p. 45

Mazzei, P.,  De Zotti, G., Xu, C.,: 1994, Ap. J. 422, 81.

Press, W. H., Flannery, B.P., Teukolsky, S. A., Vetterling, W. T. 1990,
''Numerical recipes''
Cambridge University Press , pag 203.

Rubin, V. C., Ford, W.K., Thonnard, N.,  1980, Ap J 238, 471

Salpeter, E.E.: 1955,  A. A. 94, 175.

Schmidt, M.: 1959, Ap. J. 129, 243.

Schultz, A.B., Spight, L.D., Rodrigue, M., Colegrove, P.T. \& DiSanti, M.A.
: 1991, BAAS, 23 953

Schweitzer, F.: 1990, in '' Dynamic and Interactions of Galaxies'', Proc. of
the
Int. Conf. Heidelberg (1989), p. 60, ed. Wielen R., Springer-Verlag

Soifer, B.T., Houck, J.R. \& Neugebauer, G.: 1987, Ann. Rev. Astron. Astrophys.
25, 187.

Struck--Marcell, C. \& Higdon, J.L 1993, Ap. J. 411, 108

Theys, J. C., Spiegel E. A. : 1977, Ap. J. 212,  616

Toomre A., in "The large scale structure of the Universe" , IAU Symp 1978
Longair and Einasto eds. , p. 109-116

Weil, M. L., \& Hernquist, L. 1993, AP. J. 405, 142

Wright, G.S., Joseph, R.D., Robertson, N.A., James, P.A., Meikle, W.P.S., 1988,
Mon. Not. R. astr. Soc. 233, 1.

Xu, C., De~Zotti, G.; 1989, A.A. 225, 12.

\vfill

\section{Figure captions}

{\bf Fig. 1:} R isophotes of our sample of ring galaxies.
\medskip

{\bf Fig. 2a:} shows the time evolution of an edge--on system
generated by an intrusion with an impact velocity of $160\, km\,s^{-1}$.
The radius of the intruder, a King sphere, is $0.28$ of the target disk.
The time step between different panels is  $ 0.074\,$ Gyr.
\medskip

{\bf Fig. 2b:} is a face--on view of the time evolution for the same system
as in panel a.
\medskip

{\bf Fig. 3:} shows the time evolution of a system generating
a ring galaxy endowed with a nucleus.
The time step between different panels is $ 0.03\,$ Gyr.
\medskip

{\bf Fig. 4:} shows the time evolution of a system generating
an empty ring galaxy.
The time step between different panels is  $0.03\,$ Gyr.
\medskip

{\bf Fig. 5:} panels a), b) and c) show the time evolution of rotation velocity
and its polar and radial component, respectively, for the same model as in
Fig. 2 (upper panels).
The time step between different panels is $ 0.074\,$ Gyr.
\medskip

{\bf Fig. 6:} presents the behaviour of the surface density (averaged on
concentric anuli) of the simulated system at different times; panel a):
intrusion within the first regime of velocities;  panel b): intrusion within
the
second regime.
\medskip

{\bf Fig. 7:} shows the time evolution of the gas fraction, $f_g$, gas
metallicity, $Z/Z_{\odot}$, and optical depth, $\tau/\tau_{disk}$, where
$Z_{\odot}$ is the solar metallicity and $\tau_{disk}$ the effective optical
depth for a pure disk (MXD92), for models computed with different $m_l$ values
and with a burst of intensity $i=60$ and $i=300$ lasting $0.35$ and $0.20\,$Gyr
respectively.
\medskip

{\bf Fig. 8:} shows the time evolution of the observed colors, B-V and V-K for
the same models as in Fig. 7. During the burst the time step for the plot is
0.01 Gyr.
\medskip

{\bf Fig. 9:} pictures (a-e) compare the predicted SEDs with observational data
(filled squares) for VV789, VV330, VV787,  VV32  and NGC 2973 respectively.
Heavy lines correspond to "$I_{r_0}$ case", light lines to "$I_0$ one"  (see
text); in picture (f) the former SEDs, long-short dashed line for VV32,
short-dashed line for VV787, dotted short--dashed VV330, long--dashed VV789,
dotted long-dashed NGC 2973, are compared  with M82 data, dots (UBV fluxes come
from RC3 catalogue, the other ones from Klein et al. 1988). Curves are
normalized to B band.
\medskip

{\bf Fig. 10:} pictures (a-e) compare the SEDs matching the burst (heavy lines)
with those expected at an age of $14\,$Gyr (light lines); heavy curves refer to
"$I_{r_0}$ case". The postburst evolution (see text) has been computed using
$R_{w/c}=0.43$, $I_0=4I_{loc}$ for VV789, VV330, VV787, $I_0=7I_{loc}$ for VV32
and $I_0=2I_{loc}$ for NGC 2973; the length of the burst is 0.35 Gyr with the
exception of VV32 where a burst lasting 0.2 Gyr  has been assumed, and VV 787
with 0.15 Gyr.

In picture (f) the postburst SEDs (symbols are as in Fig. 9) are compared with
that of our galaxy (dotted line) at the same age (MXD92). Curves are normalized
to B band.
\medskip

{\bf Fig. 11}: panel a) shows our morphological fit for VV789, the isodensity
contours are taken to a stage $\tau=0.20$ from the beginning of the ring
structure; panel b) refers to VV330 with $\tau=0.40$; panel c): VV32, t=0.044
Gyr from the beginning of the ring structure; panel d):  NGC 2793, $\tau= 0.17$
\vfill\eject
\end{document}